\documentclass[preprint,12pt]{aastex}

\shorttitle{MIRI coronagraphs}
\shortauthors{Boccaletti et al.}

\begin{document}

\title{The Mid-Infrared Instrument for the James Webb Space Telescope, V: Predicted Performance of the MIRI Coronagraphs}

%
%
%
%
%


\author{A. Boccaletti\altaffilmark{1}, P.-O. Lagage\altaffilmark{2}, P. Baudoz\altaffilmark{1}, C. Beichman\altaffilmark{3.4.5},  P. Bouchet\altaffilmark{2}, C. Cavarroc\altaffilmark{6},  D. Dubreuil\altaffilmark{2}, Alistair Glasse\altaffilmark{7}, A. M. Glauser\altaffilmark{8}, D. C. Hines\altaffilmark{9}, C.-P. Lajoie\altaffilmark{9}, J. Lebreton\altaffilmark{3,4}, M. D. Perrin\altaffilmark{9},  L. Pueyo\altaffilmark{9}, J. M. Reess\altaffilmark{1}, G. H. Rieke\altaffilmark{10}, S. Ronayette\altaffilmark{2}, D. Rouan\altaffilmark{1},  R. Soummer\altaffilmark{9},  and G. S. Wright\altaffilmark{7} }

\altaffiltext{1}{LESIA, Observatoire de Paris-Meudon, CNRS, Universit\'e Pierre et Marie Curie, Universit\'e Paris Diderot, 5 Place Jules Janssen, F-92195 Meudon, France}
\altaffiltext{2}{Laboratoire AIM Paris-Saclay, CEA-IRFU/SAp, CNRS, Universit\'e Paris Diderot, F-91191 Gif-sur-Yvette, France}
\altaffiltext{3}{Infrared Processing and Analysis Center, California Institute of Technology, Pasadena, CA 91125, USA }
\altaffiltext{4}{NASA Exoplanet Science Institute, California Institute of Technology, 770 S. Wilson Ave., Pasadena, CA 91125, USA}
\altaffiltext{5}{Jet Propulsion Laboratory, California Institute of Technology, 4800 Oak Grove Dr., Pasadena, CA 91107, USA} 
\altaffiltext{6}{Laboratoire AIM Paris-Saclay, CEA-IRFU/SAp, CNRS, Universit\'e Paris Diderot, F-91191 Gif-sur-Yvette, France}
\altaffiltext{7}{UK Astronomy Technology Centre, STFC, Royal Observatory Edinburgh, Blackford Hill, Edinburgh EH9 3HJ, UK }
\altaffiltext{8}{ETH Zurich, Institute for Astronomy, Wolfgang-Pauli-Str. 27, CH-8093 Zurich, Switzerland}
\altaffiltext{9}{Space Telescope Science Institute, 3700 San Martin Drive, Baltimore, MD 21218}
\altaffiltext{10}{Steward Observatory, The University of Arizona, 933 N. Cherry Ave., Tucson, AZ 85721, USA}

\begin{abstract}
The imaging channel on the Mid-Infrared Instrument (MIRI) is equipped with four coronagraphs that provide high contrast imaging capabilities for studying faint point sources and extended emission that would otherwise be overwhelmed by a bright point-source in its vicinity.  Such bright sources might include stars that are orbited by exoplanets and circumstellar material, mass-loss envelopes around post-main-sequence stars, the near-nuclear environments in active galaxies, and the host galaxies of distant quasars. This paper describes the coronagraphic observing modes of MIRI, as well as performance estimates based on measurements of the MIRI flight model during cryo-vacuum testing.  A brief outline of coronagraphic operations is also provided. Finally, simulated MIRI coronagraphic observations of a few astronomical targets are presented for illustration.

\end{abstract}


\section{Introduction}

High contrast imaging has made major strides since the first coronagraphs were used by Lyot to investigate the Solar Corona \citep{lyot1939}.  The success of coronagraphy using the {\it Hubble Space Telescope (HST)}\footnote{The Near Infrared Camera and Multi-Object Spectrometer (NICMOS), Space Telescope Imaging Spectrograph (STIS), and the Advanced Camera for Surveys (ACS) all provided coronagraphic capabilities.} showed the clear advantage of the very stable environment achieved with space-based platforms, while the complementary successes of ground-based systems showed the importance of achieving very small Inner Working Angles (IWAs).

The {\it James Webb Space Telescope (JWST)} was recognized during its early development as a powerful platform for studying exoplanets, or more generally, circumstellar environments. In consequence, coronagraphic devices were foreseen as integral components of the science instruments. The potential is particularly large in the deep thermal regime where  the Mid-Infrared Instrument ({\it MIRI}) operates, wavelengths where very faint objects can be detected by {\it JWST} in comparison with the limited sensitivity of even the largest ground-based telescopes. Therefore, in addition to a classical Lyot coronagraph, {\it MIRI} incorporates one of the most innovative concepts, the four-quadrant phase mask coronagraph \citep[4QPM,][]{rouan2000} to achieve this capability. At the time of the MIRI conceptual design phase, only a few coronagraphic technologies had been validated both in the laboratory and on-sky \citep{baudoz2006, boccaletti2004}; the 4QPM appeared to be the best option to provide the smallest possible IWA of $\sim 1\lambda/D$ at MIRI wavelengths (in this case, 10 - 16 $\mu$m). 

Direct imaging of exoplanets is one of the most challenging topics in modern astrophysics as it requires a large instrumental contrast, and hence very good optical quality. By the time of {\it JWST's} scheduled launch in late 2018, SPHERE 
 \citep[Spectro-Polarimetric High-contrast Exoplanet Research,][]{beuzit2008} and GPI \citep[Gemini Planet Imager,][]{macintosh2008}
will have been operating for nearly five years. These near-IR (0.95-2.3$\mu m$), eXtreme Adaptive Optics (XAO) facilities will have likely discovered dozens of young giant planets that {\it JWST} will be able to observe at complementary wavelengths (5-28$\mu m$), improving the spectral characterization of the planet atmospheres and measuring their energy balance. In addition, the Large Binocular Telescope Interferometer (LBTI) operating in the thermal infrared \citep{hinz2009} will provide insights to the behavior 
of planetary systems at higher angular resolution than can be achieved with MIRI on JWST. 

In addition, MIRI coronagraphy will enable a range of investigations to measure structures in circumstellar disks, potentially identifying locations of forming planets, or the influence of unseen planets on the circumstellar material.  It will also enable detailed studies of the circumnuclear environments in active galaxies, potentially probing the connection between star-formation and outflows powered by the central supermassive black holes.  Thus, high contrast capabilities provided by MIRI in the mid-infrared will open new frontiers in a variety of astrophysical contexts that have never been accessible before.

This paper updates the descriptions of the MIRI coronagraph performance in \citet{boccaletti2005} and \citet{cavarroc2008}. 
Section \ref{sec:modes} introduces the coronagraphic modes available in MIRI and section \ref{sec:perf} describes the numerical simulations and the expected raw contrast performance.
Section \ref{sec:labtest} presents the contrast performance measured in the lab during the integration and test of the flight model.
Section \ref{sec:TA} discusses the target acquisition procedure that is specific for the MIRI coronagraph. 
 The potential interest of MIRI for exoplanet and disk science is illustrated in section \ref{sec:illustration} together with simulated observations of known targets. 
Finally, we indicate in section \ref{sec:futurecapab} some possible expansions of the high-contrast capabilities of MIRI.

\section{Coronagraph Implementation}
\label{sec:modes}
MIRI has four individual coronagraphs, one of which is based on the classic design of Lyot and three of which are based on four-quadrant phase masks (4QPMs).  The classical Lyot coronagraph places an occulting spot in the focal plane to block the light from a bright point source from entering the instrument. 
Typically, this spot is of order $3-6\lambda/D$ in radius so that it blocks the majority of the light from the Airy core including a few bright rings.
Nonetheless, some extraneous light escapes 1.) as the diffraction pattern associated with the telescope aperture and a particular coronagraph design, and 2.) as phase and amplitude aberrations of the wavefront in the optical train.
Some of the unwanted light can be intercepted by appropriate Lyot stops, at the expense of lost light, at a reimaged pupil. These stops can remove the diffracted light induced both by the telescope pupil and the coronagraphic occulting spot\footnote{In a coronagraph free of optical aberrations, the distribution of the intensity in the reimaged pupil is determined by the convolution of the telescope pupil and the Fourier Transform of the coronagraphic mask.}. 
However, completely suppressing the telescope diffraction pattern resulting from the multi-mirror {\it JWST} primary would require relatively complex stops.
For MIRI, a Lyot coronagraph is provided at 23 $\mu$m, selected in part because of the lack of suitable transmissive optical materials to fabricate a 4QPM and in part to provide a broad spectral band to maximize the sensitivity on planetary debris disks (currently available 4QPMs are not achromatic). 

Although classical Lyot coronagraphs can provide excellent contrast outside the area blocked by the occulting spot, obviously they are limited with regard to IWA to the projected radius of this spot ($\ge 3 \lambda/D$). 
Phase mask coronagraphs are designed to reduce the IWA to near $1 \lambda/D$, by replacing the occulting spot with a transparent mask that imparts phase differences in different parts of the
focal plane so when the reimaged pupil is formed the light interferes more destructively than with a Lyot mask, hence rejecting the starlight outside the geometrical pupil. 
There are several implementations of phase-mask coronagraphs based on this principle. The theory of 4QPMs is developed in \citet{rouan2000} and \citet{rouan2007}.

\begin{figure}[htbp]
\centerline{\includegraphics[width=6.0in]{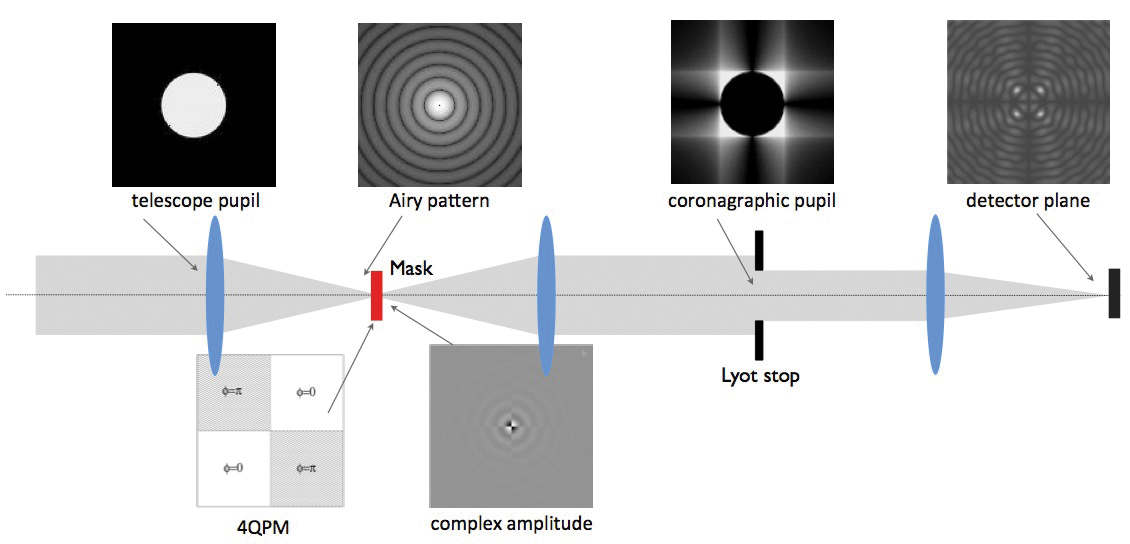}}
\caption{Principle of operation of the 4QPM coronagraph. The telescope is represented by the large lens to the left, for simplicity assumed to be illuminated uniformly and to be round 
as shown by the circular pupil sketch to the lower left. It forms a point source image on the 4QPM as an Airy pattern. The phase plate modifies the wavefronts to produce the
complex amplitude pattern indicated for a source centered exactly on the intersection of the quadrants.  The lens behind the telescope focal plane forms a pupil where a traditional
Lyot stop is placed to remove part of the diffracted light from the telescope pupil through the coronagraphic mask.
When the following lens reimages the focal plane onto the detector, the complex amplitude pattern causes the light from the bright central source to interfere destructively. Figure adapted from \citet{rouan2007}. The layout of a Lyot coronagraph is identical except for the substitution of an occulting spot for the 4QPM.}
\label{fig:theory}
\end{figure}

 The 4QPMs in MIRI use an optical element that retards the phase by $\pi$ in two diagonally opposite quadrants. If a monochromatic source is placed 
exactly at the center of the resulting four-quadrant phase mask, the rejection is formally complete. The overall operation is illustrated in Figure \ref{fig:theory}. 
The MIRI phase masks are monochromatic versions of the 4QPM, that is they are effective at a single wavelength. The MIRI devices generally operate with spectral bandpasses of about 10\%. However, the net coronagraphic rejection suffers from several limitations including phase and amplitude aberrations, which are larger in the case of MIRI than the chromaticity caused by the filter bandwidth \citep{riaud2003}. 
As another defect, a point-source lying in between two quadrants is also partly attenuated \citep{riaud2001}. Therefore, the 4QPM reduces the sensitivity (at maximum by a factor of 10) in the field along the four edges of the mask (in a width of about 1$\lambda$/D).

\begin{figure}[htbp]
\centerline{\includegraphics[width=5.0in, height=3.0 in]{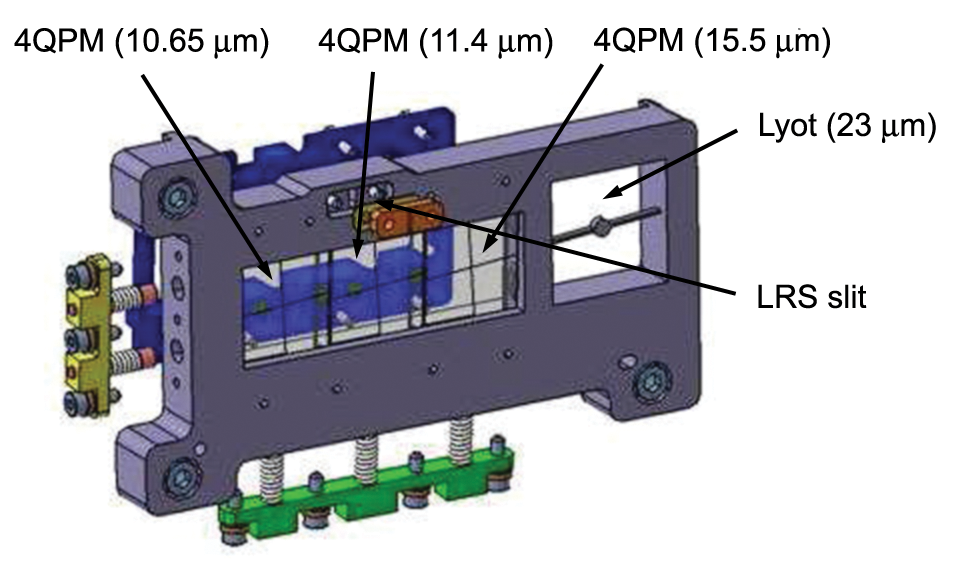}}
\caption{Mounting bracket for the four coronagraph image-plane masks and the LRS slit. The action of the spring-loaded mounting hardware is shown as an exploded view.}
\label{fig:mount}
\end{figure}

\begin{figure}[t]
\centerline{
\includegraphics[width=12cm]{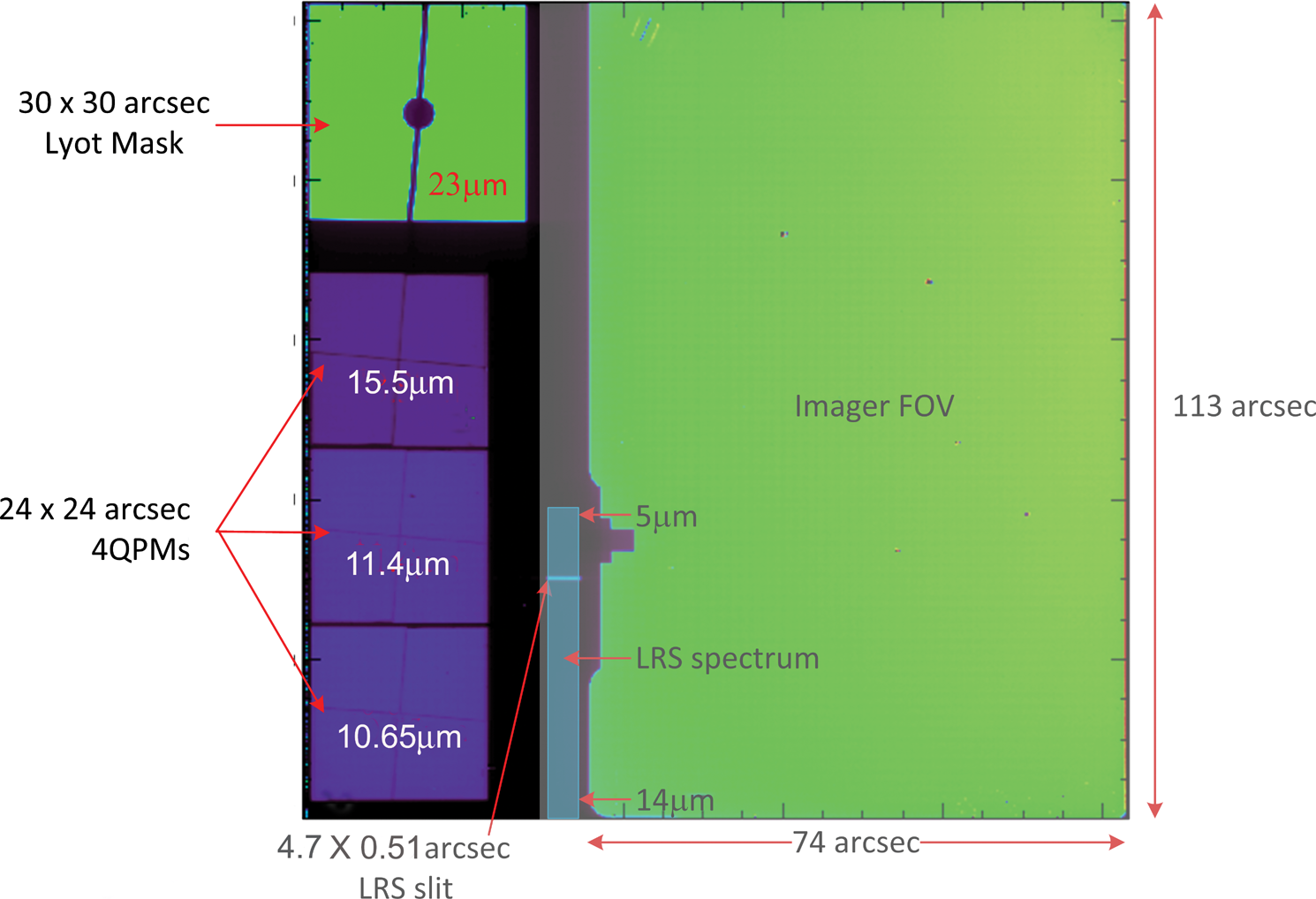}}
\caption{Flat field image: the various coronagraphic fields are located to the left (phase masks at the bottom, and the Lyot mask at the top), while the imager field is at the right.}
\label{fig:flat}
\end{figure}

All four of the coronagraphic masks are implemented in the entrance field of the MIRI imager on a frame supporting four sub-fields of view; the three 4QPMs provide fields of $24^{\prime\prime} \times 24^{\prime\prime}$, while the free-standing Lyot spot mask is held by supporting bars across a $30^{\prime\prime} \times 30^{\prime\prime}$ aperture. 
The mounting arrangement,  along with that for the low resolution spectrograph slit, is shown in Figure \ref{fig:mount}. 
The location of the coronagraphic fields is shown in a flat field image (see Figure \ref{fig:flat}); they occupy about a third of the detector square field of view. 
No mechanism is required to utilize the coronagraphs - the telescope is used to position a star on each one as required. 

The Lyot spot mask is machined in aluminium and has a radius of $2.16^{\prime\prime}$, which is 3$\lambda$/D in radius at 23$\mu m$ 
The 4QPMs are etched in germanium (Figure 4) and then anti-reflection coated. The pupil masks, mounted in the filter wheel, are shown in Figure \ref{fig:pupil}. The four masks are each associated with an appropriate filter (labeled "C" in Table \ref{tab:filters}) 
together with an appropriate Lyot stop in the filter wheel where the pupil image is formed. The stops have a transmission of 62\% and 72\% for the 4QPMs and the Lyot spot mask, respectively. 
The phase masks were designed to operate at 10.65, 11.40 and 15.50 $\mu m$ for exoplanet science while the Lyot coronagraph is optimized for 23 $\mu m$, in particular for disk imaging. 
The two bands near 11 $\mu$m were selected to allow detection of the NH$_3$ feature at 10.65 $\mu$m in exoplanetary atmospheres.  
The choice of filter wavelengths for MIRI coronagraphs is discussed further in \citet{boccaletti2005}.
In practice, the central wavelengths of manufactured filters measured at nominal temperature are well controlled (1\% offset at maximum), while the bandwidths can be broader by 40\% at maximum. 

A summary of coronagraphic configurations is given in Table \ref{tab:filters}.
The rejection is calculated on-axis and accounts for all the aberrations considered in the simulation (see section \ref{sec:perf}).
The coronagraphs produce a radial attenuation plotted in  Figure \ref{fig:iwa}, which quantifies the total attenuation of a point-source located on-axis. The 50\% transmission radius, defined as the IWA, is about  1$\lambda$/D for the 4QPMs and 3.3$\lambda$/D for the Lyot spot mask. 100\% transmission (relative) is achieved at 3 and 5$\lambda$/D for respectively the 4QPM and the Lyot spot.

\section{Laboratory tests}
\label{sec:labtest}

\subsection{Prototype units}
4QPM coronagraphs were tested in the laboratory by \citet{riaud2003} and  demonstrated on the telescope \citep{boccaletti2004, 
boccaletti2009, boccaletti2012, gratadour2005, riaud2006}. The laboratory results showed very high performance (raw contrast approaching 10$^6$), although use on ground-based telescopes was challenging because of the level of aberrations
left uncorrected by adaptive optics as well as the need to maintain the star precisely on the
center of the phase mask, a near-corollary of the small IWA. 

This issue will be mitigated
by operation in space with JWST.  Therefore, the focus has shifted to tests of prototypes of the 4QPM coronagraphs for MIRI as described by \citet{baudoz2006} and \citet{cavarroc2008}. 
The first of these papers concentrates on the specifications and manufacture of the 4QPM itself, optimized for the mid-infrared as required for MIRI. It considers the requirements on the chromaticity of the mask material, how well its dimensions must be matched to the wavelength of operation, and the fidelity that must be achieved at the intersection of the four quadrants (where the star image must be placed). It shows the results of a variety of manufacturing possibilities and demonstrates good results using germanium as the mask material. The steps between quadrants on this successful mask were reactive ion etched. The second paper is concerned with the operation of 4QPMs specifically in MIRI. It describes measurements of the 
performance of pre-flight phase masks in the MIRI imager 
test system at CEA-Sacley, including practical effects such as a mis-alignment of the instrument pupil with the telescope exit pupil. These experiments led to further optimization of the coronagraph. The performance with the flight units (next section) was predicted accurately in this work. 

\subsection{Verification in flight instrument}

The performance of the flight coronagraphs was verified in the flight model imager prior to its delivery to be integrated with the full MIRI Optics Module (OM) \citep{ronayette2010}. The test arrangement of a warm source outside the cryostat with heavy filtering to block room-temperature thermal radiation is described in Paper III. A 30 $\mu$m diameter pinhole source (much smaller than the FWHM of the diffraction pattern) in the external apparatus was scanned in a raster pattern to measure the response of the coronagraphs, with results shown in Figure \ref{fig:fmimage}. The faint extended residual in the image to the right
is due to local heating of the surroundings of the pinhole infrared source, which is not rejected
by the coronagraph and places an artificial floor on its contrast performance. 

The coronagraph shows good rejection of the central point spread function (PSF), by a factor larger than 100 on-axis (up to a few hundreds), 
although the level of rejection is probably limited more by the experimental limitations than by the coronagraph.  
The scans also showed that the IWA, defined as the angle with 50\% of the maximum transmission, is at 1.3
$\lambda$/D for the 4QPM units. Figure \ref{fig:fmcontrast} shows the measured contrast as a function of distance from the center of the PSF, and demonstrates that it agrees closely with theoretical expectations.

\section{Simulation and Performance}
\label{sec:perf}

\subsection{Coronagraph contrast performance}

With the performance of the coronagraph itself confirmed, we now turn to simulations of its operation in JWST. Two models are under evaluation: Case A makes relatively optimistic assumptions about the wave front error, while Case B is more conservative. 

The simulation of MIRI coronagraphy under Case A is performed in two steps, first the calculation of coronagraphic images and PSFs themselves and then a photometric treatment that includes sources of noise. The images are modeled with a Fourier-based approach described in \citet{boccaletti2005}.\footnote{The simulation software was used formerly to derive specifications of the MIRI coronagraphs and to estimate performance.} The model accounts for the telescope pupil (18 hexagonal segments separated by small gaps) and phase aberrations (independent on each segment), for a total of  $\sim$130\,nm root mean square (rms) (mid and high frequencies) wavefront error (WFE), as well as a global low order aberration in the form of defocus equivalent to 2\,mm at the telescope focal plane. In the pupil plane (filter wheel), the position of the Lyot stop can be mis-aligned so it is offset by 3\% of the size of the pupil and includes also a 0.5$^{\circ}$ rotation with respect to the perfect pupil orientation. To model the telescope pointing variations the calculation of the coronagraphic image is repeated 200 times with a 1-sigma dispersion of 7\,mas  (corresponding to the current specifications for JWST) and all realizations are summed in the end. A reference star image is calculated simultaneously (same setup as the target) but considering different jitter realizations. We assume that the target/reference duty cycle can be short enough to neglect the variation of the wavefront errors and consider that the target to reference mismatch is driven by the pointing precision of 5\,mas (the {\it JWST} pointing requirement at 1-sigma/axis). Finally, several off-axis images (PSFs) are generated in the field at various positions ($0.1^{\prime\prime}, 0.2^{\prime\prime}, 0.5^{\prime\prime}, 1^{\prime\prime}, 10^{\prime\prime}$) to account for the coronagraphic radial attenuation (see  Figure \ref{fig:iwa}).  Both types of coronagraphs are included in the simulation, the 4QPM and the Lyot spot mask. For the broad band filter F2300C, polychromatic images are generated according to the tabulated transmission of this filter; the 4QPM coronagraphs were simulated with their integral filters. Raw coronagraphic contrasts (azimuthally averaged), corresponding to the first step of the simulation (free of detection noise), are shown in  Figure \ref{fig:rawcontrast}, together with 5-$\sigma$ post-processing contrasts after reference star subtraction. This last step provides an estimate of the maximum contrast achievable directly with this technique. However, in addition to this simple, single reference star subtraction, several more advanced post-processing techniques  have been developed \citep{lafreniere2007, soummer2012} and should be implemented to improve this limit of detection. 

We now consider Case B, which assumes a WFE of 204 nm and the specification offsetting accuracy (dispersion of 7 mas 1-$\sigma$ per axis with different realization for the star and the reference). Jitter is not included. Model results are shown in Figure \ref{fig:rawcontrast}. 
While the achieved contrast is a fairly strong function of optimal positioning within the coronagraph, it is even more sensitive to the relative locations of the science target and the reference star \citep[see, e.g, Fig. 1:][]{cavarroc2008}. The comparison with the Case A simulation reflects our current uncertainty about the exact parameters that are applicable and how their effects change with wavelength (e.g., the current work indicates a greater difference between Cases A and B at 15.5 $\mu$m than at 11.4 $\mu$m). Work is continuing to understand these differences and to use the knowledge gained to optimize coronagraphic observations.    

\subsection{Full system performance}

The second part of the simulation includes the detection noise. We illustrate the procedure for Case A. The photon noise in the coronagraphic image depends on the star distance and spectral type, telescope area (25\,$m^2$), telescope and optics transmission (85\%), the detector spectral response (i.e., quantum efficiency maximum of $\sim$80\% at $\sim$15$\mu m$ ), the Lyot stop transmission (62\% for 4QPM, 72\% for Lyot spot and 100\% for Lyot bar), and the integration time. The net transmission in the coronagraphic filters is about 15-20\%. The intensity of the most distant off-axis PSF at 10$''$ serves as a normalization factor for all simulated images, as if it were an observation of the unocculted star.  In addition, we assume that the reference star is identical to the target star in terms of magnitude and spectrum. 

At the level of the detector, we included read out noise (20e- RMS), and flat field stability (0.1\%).
The thermal background accounts for several components: the zodiacal light dominates at $\lambda<15\mu m$, while at longer wavelengths the telescope (primary and secondary mirrors) as well as the deployable tower assembly shield  and sunshield are mostly responsible for the thermal emission. Overall the model corresponds well to the specifications of  3.9\,MJy/sr at 10$\mu m$  and 200\,MJy/sr at $20\mu m$ (the possible higher emission from the sunshield than originally expected will compromise the performance of the Lyot system, but has less effect on the shorter wavelength coronagraphs).

Figure \ref{fig:perf} displays four cases for various filters. Each sub-panel shows the averaged contrast for the PSF (blue line) and the coronagraphic image (red solid line) as well as the 3-$\sigma$ contrast for the reference star subtraction process (red dashed line) and the ideal noise level if only photon noise dominates (red dash-dotted line).
Typical contrasts after post-processing achieve $10^{-4}$ to $10^{-5}$ for separations larger than $0.5^{\prime\prime}-1^{\prime\prime}$. 
Note that the radial transmissions of the MIRI coronagraphs are not accounted for in these contrast profiles.

The performance presented in this section is strongly related to the end-to-end optical system assumptions, such as the wavefront error and the whole optical train stability of the telescope. The simulated performance may be conservative, particularly with the development of operations and analysis techniques optimized for JWST. In any case, the current model will have to be updated once the telescope is on orbit and commissioned. 


\section{Operations and Target Acquisition}
\label{sec:TA}
MIRI operations, including coronagraphy, are discussed in Gordon et al. (2014: Paper X).  Here we briefly discuss the operations unique for coronagraphy, particularly target acquisition (TA). To achieve maximum contrast, all coronagraph observations will require TA to center the target accurately on the coronagraph.  The 4QPMs are particularly sensitive to slight offsets from the optimum location \citep[Figure 1:][]{cavarroc2008}. The Lyot coronagraph is less sensitive to small offsets, but still requires precise, and repeatable TAs.   

\subsection{Target acquisition}

The basic approach to coronagraphic TA starts by selecting a suitable filter.  For faint objects, the filter can simply be that associated with the coronagraphic mask, but for very bright objects, a neutral density filter will be needed.  The target is initially placed at a fiducial location within the coronagraphic field of view. An exposure is obtained, a centroid is found for the target, and the offset necessary to move the target to the optimal location at the center of the coronagraph is calculated. The observatory then makes a small angle maneuver to place the target at the proper location in the coronagraph. Once the maneuver is complete the observatory enters ``fine lock,'' which maintains precise pointing control throughout the observation. If the neutral density filter was used for TA, then the correct mask filter is selected for observations (the ordering of this step is under study, in case it is advantageous to move to the mask filter before maneuvering to the coronagraph position). 
A TA must be executed for observations obtained with each coronagraph separately.  Thus, a science program that requires observations using multiple coronagraphs will need to take these additional TAs into account.

Two effects make the TA process complex: 1.) for the 4QPM coronagraphs, the phase mask can distort the image of a star close to its center and undermine the accuracy of the centroid determination; and 2.) the detector arrays have latent images (Paper VII) that could mimic planets or other exciting astronomical phenomena if the centroiding process left them close to the target star. These effects would make adequate TA very difficult at the nominal JWST offsetting accuracy specifications. Fortunately, it is projected that small angle offsets up to 20$''$ distance will be much more accurate than dictated by the requirements. For example, an offset of up to 20$''$ is expected to be accurate to 5 mas (1-$\sigma$ per axis). Simulations of the centering accuracy on the coronagraph using the projected performance and a fiducial distance of 2$''$ from the coronagraph center indicate a scatter of $\sim$7 mas (rms) and average centering errors of 2 - 4 mas \citep{lajoie2014}. The details depend on the particular strategy, i.e., whether one utilizes a single position for target acquisition, or uses more than one to acquire additional information about the pointing. None of the strategies quite reaches the desired centering performance for the 4QPM coronagraphs (the Lyot is much more relaxed in this area), so further optimization is in progress.   

\subsection{Coronagraphic observations and results}

After the TA process is complete, images are acquired until the necessary on-source integration time is achieved.  The details of the actual operation of the detectors, their calibration and data reduction are described 
in Ressler et al. (2014, Paper VIII) and Gordon et al. (2014, Paper X).

To maximize contrast, most coronagraphic observations will require coronagraphic observations of a ``reference star,'' whose image is then subtracted from the science target image(s). Typically this will be a star that is of similar or greater brightness compared with the science target; this ensures equal or greater signal to noise for the reference star in a similar or shorter amount of total integration time.  The procedure for obtaining these observations is identical to that for the science target, including the TA.


As mentioned in Section 4, several techniques and algorithms for improving contrast have been developed that are more sophisticated than the classical reference star subtraction method.  One is angular differential imaging (ADI), where the target and reference star are observed at multiple orientations of the observatory.  Because the diffraction patterns and detector artifacts are fixed to the telescope reference frame, they move relative to the target when the observatory is rolled. Subtracting two (or more) rolled images removes, or at least reduces, these diffraction and detector features.  {\it JWST} has limited roll capability ($\pm 5^{\circ}$), because it must keep the sunshield correctly positioned relative to the Sun, but will still provide sufficient angular roll for some science cases at large enough angular separations from the star.  Larger roll angles can be achieved by observing the objects at different epochs during the year albeit with more significant changes in the wavefront error compared to back-to-back observations. Another class of enhancement is centered on sophisticated image
analysis techniques, such as LOCI or KLIP \citep{lafreniere2007, soummer2012}.


\section{Illustration of Dedicated Science Cases}
\label{sec:illustration}

This section presents a few science cases making use of the capabilities of the MIRI coronagraphs.  We compare the simulated performance metrics presented in Section \ref{sec:perf} with the expected flux densities from giant planets orbiting solar type stars out to 10 pc.  We also present simulated observations of the emblematic HR\,8799 planetary system discovered by \citet{marois2008}, and observations of the debris disk associated with HD\,181327. These examples illustrate the promise of the MIRI coronagraph in the context of exoplanet and disk science. In the case of exoplanets, MIRI coronagraphic imaging will: 1.) measure effective temperatures;
2.) determine bolometric luminosities; 3.) measure the ammonia absorption feature (sensitive to the presence of ammonia clouds and to the atmospheric temperature); and 4.) in combination with shorter wavelength measurements, support comprehensive modeling of the atmospheric properties \citep[e.g.,][]{bonnefoy2013}. In the case of debris disks,  the capability will let us: 1.) explore the virtually unknown region between about 1 and 30 AU, containing for example the ice lines (not resolved with {\it Spitzer}, too low in surface brightness for HST or groundbased telescopes); 2.) constrain the minerology of planetary debris; and 3.) search for structures maintained by larger bodies in planetary systems.

\subsection{Exoplanet imaging}

Colored symbols in  Figure \ref{fig:perf} stand for the intensity of Jovian planets orbiting G and M stars at 10 pc and for different assumed planetary temperatures. The radial transmission of the coronagraph for each modeled planet is highlighted with dotted lines.

For synthetic photometry of planets, we used the BDcond2000 models of \citet{allard2001} integrated over the MIRI filters. Several test cases were considered: 300\,K, 400\,K, 500\,K with a radius of 1.0\,R$_{ J}$ and 500\,K, 1000\,K, 1500\,K with a radius of 1.4\,R$_{ J}$.

At a separation of $1^{\prime\prime}$, the contrasts expected, even for simple reference star-subtracted coronagraphy, enable detection in F1140C of 1.0\,R$_{ J}$ planets with temperature of 500\,K (respectively 350\,K) orbiting a G0V(respectively M0V) star. The situation is more favorable in F1550C where a 400\,K temperature is reached at $1^{\prime\prime}$ from a G0V star.  For example, the $\beta$ Pic b planet \citep{lagrange2010} has a temperature of 1600-1700\,K and a large radius \citep[1.3-1.6\,R$_{ J}$,][]{bonnefoy2013}, hence should be brighter than any planet in  Figure \ref{fig:perf}, so almost visible in raw coronagraphic images (providing the debris disk itself is fainter than the planet).  Post-processing should yield a very high signal to noise detection.
 
\subsection{Imaging HR 8799}

For the simulations of HR\,8799, we assumed an A0V star at 40pc, and 3 hours of on-source integration in each MIRI filter. The planets b, c, and d, were assigned temperatures of 900\,K, 1100\,K and 1100\,K with identical radii of 1.25\,R$_{J}$. The corresponding contrasts in each filter, according to \cite{allard2001}, are listed in Table \ref{tab:hr8799}. The angular separations are $1.72^{\prime\prime}, 0.96^{\prime\prime}, 0.62^{\prime\prime}$ for planet b, c and d. Some of the planets are close enough to the center of the 4QPM coronagraphs to be attenuated significantly. Planet b is always at a separation larger than 3$\lambda$/D in any filter so it is not impacted. Planet c is inward of 3$\lambda$/D in F1065C, F1140C and F1550C, and hence is attenuated by 10, 15 and  30\%, respectively. Similarly,  planet d is attenuated by 
35, 40, 40\% in the same coronagraphic filters. Simulated images are displayed in  Figure \ref{fig:imageshr8799}.

Planets b and c are detected at all wavelengths, while planet d is only visible at 15.5$\mu m$. Signal to noise ratios measured in the brightest pixel are given in Table \ref{tab:hr8799snr}. These numbers, if integrated over a PSF size can be as large as 100 for planet b, for instance. The apparent elongation of planet d in the F1550C filter (in which the star residuals are fainter than the planets and the background level) is in fact due to the presence of planet e nearby. These images were used to measure the planet photometry and to compare with the initial planet intensity. An example is shown in  Figure \ref{fig:spectrumhr8799} for planet b, where the solid line represents the model degraded to a spectral resolution of $R=20$, red points stand for the initial photometry and blue points are the measurements obtained in each filter. There is good agreement in the four filters around $\lambda=10\mu m$. 
At  F1550C the departure of the measured intensity from the true value is larger and likely the result of residual diffraction on one side and background noise on the other side. This does not mean MIRI will fail to obtain accurate photometry in these cases but rather that a special analysis will be required. 

\subsection{The HD 181327 debris disk}

Another example of MIRI coronagraphic capabilities is given for the case of an extended object. Figure \ref{fig:hd181327} shows simulated images of the ring-like debris disk around the star HD\,181327 \citep{schneider2006}. Here, we assumed an F5V central star, located at 50pc, and observed for 1 hour in each MIRI filter. 
The disk model corresponds to a $\sim$90 AU-wide belt containing 0.05 $M_{\oplus}$ 
(up to 1 mm) of icy and porous, silicates and carbonaceous dust grains in collisional equilibrium. This model was found by \citet{lebreton2012}  to be  the best Bayesian solution to reproduce the disk Spectral Energy Distribution (SED) from mid-infrared to millimeter wavelengths, assuming the radial brightness profile measured in newly-reduced {\it HST}/NICMOS scattered-light images. Scattering and absorption efficiencies as well as equilibrium temperatures were computed as a function of grain size distribution and composition using state-of-the-art radiative transfer techniques. Monochromatic images were produced at the effective wavelength of each filter. They include a scattering phase function with $g_{HG}=0.3$ as seen in the {\it HST} images; their absolute flux is conservatively scaled to the {\it HST} one at 1.1 $\mu$m to correct for the limited capability to predict scattering properties. The transition between scattering- and thermal emission-dominated regimes occurs around 10 $\mu$m, which turns out to be the least favorable band for imaging due to an unfavorable compromise between contrast and resolution. 
The corresponding star to disk total flux ratio is 
 154, 48, 3.5, 0.16, respectively in 
 F1065C, F1140C, F1550C, F2300C filters. 
 The 86\,AU ring-like pattern is obvious in all images as well as the north/south asymmetry. The 23$\mu m$ image is sensitive to the outer part of the disk while the central cavity is blocked by the Lyot spot. Compared to the {\it HST} images obtained in the near-IR \citep{schneider2006} and visible \citep{schneider2014}, the MIRI images have greatly reduced stellar residuals in the center and will allow a cleaner analysis of the dust grain properties based on mid IR colors.  In addition, the ring disk is seen close to face-on and hence would be immune to the gain brought by angular differential imaging in ground-based observations.
 Here, for simplicity we assumed that the radial transmission of 4QPMs is isotropic, so we did not model the cross-like pattern of this coronagraph where the field is partially obscured. In practice, for extended objects, observations at multiple rolls (at least two at 45$^{\circ}$) will enable "filling in" the gaps caused by the 4QPM boundaries and by the support bar in the Lyot coronagraph.


\section{Future Capability Expansions}
\label{sec:futurecapab}

\subsection{Expanding the wavelength coverage}

MIRI coronagraphy is centered on four filters, which can be regarded as a minimum set for spectral characterization of exoplanet atmospheres considering the large spectral range available. In particular, wavelengths shorter than 10$\mu m$ carry the signatures of water vapor and methane, which are expected to be abundant in giant planet atmospheres. In principle, one can use the broad band filters for imaging (labeled "W" in Table \ref{tab:filters2}) in combination with the Lyot mask. The absence of a Lyot stop to accompany these filters significantly reduces the coronagraphic rejection. However, the bar supporting the Lyot mask is sufficiently large compared to the size of the Airy disk, especially at the shortest wavelengths, for some coronagraphic effect to occur using the bar as the focal plane occulter. At a minimum, these non-standard degraded coronagraphic modes should be considered as anti-saturation devices rather than coronagraphs, but they can still can be very beneficial for observing near bright stars. There are currently no plans to implement these modes on-orbit. However, we present them here for completeness and to advocate interest in them.

To illustrate their application, similarly to section \ref{sec:perf}, we performed simulations of the Lyot bar mask for three imaging filters (F560W, F770W, F1000W). We assume the central source is offset by 6" from the center of the field so the star falls on the bar rather than the Lyot spot. After reference star subtraction, the contrast performance is comparable to that of the coronagraphic modes ( Figure \ref{fig:perf2}) as the PSF falls off more rapidly with radius at wavelengths shorter than those passed by the coronagraphic filters. However, we note for F560W and F770W that the coronagraphic effect is really small beyond the IWA, as expected. 

\subsection{Enhancing contrast with micro-offsets}

Another possibility is to obtain multiple observations of the reference star with slightly different positional offsets.  {\it JWST} has a fine steering mirror (FSM), which is used to compensate for small perturbations in pointing control.  The FSM can also be used during coronagraphic imaging programs to perform micro-offsets (sub-pixel), thus building up a library of reference star images \citep{soummer2014} for use in the more sophisticated algorithms. Use of the FSM has the advantage that it does not require any additional TAs, both making this mode very efficient and avoiding additional errors due to small-angle maneuvers.
The image library can be used for PSF subtraction with a variety of algorithms 
(e.g; LOCI or KLIP, \citet{lafreniere2007}; \citet{soummer2012}). A preliminary evaluation (at this stage omitting
the effects of latent images and pointing jitter, both of which need to be added to future simulations) 
indicates possible gains in contrast by up to an order of magnitude, when working
close to the IWA \citep{soummer2014}. This gain comes at the expense of the increased observing time on the reference star, but could be critical to success for demanding observations.


\section{Conclusions}

The coronagraphic capability of MIRI is unique as it allows high-contrast imaging at wavelengths that are usually not considered for exoplanets and other high-contrast science. With {\it JWST} these science applications will become attractive owing to the large aperture and high sensitivity. The MIRI spatial resolution is comparable to those obtained in the near-IR in space with {\it HST}. To exploit this opportunity, MIRI includes phase mask coronagraphs, which are able to offer the same IWA as the NIRCAM coronagraphs but at longer wavelengths. Even with simple, single reference star subtraction, we have shown that the expected performance of the MIRI coronagraphs will open a new frontier in the characterization of many astronomical objects from exoplanets, to circumstellar disks, to the near-nuclear regions of nearby active galaxies and the host galaxies of distant quasars. 

For example, our performance analysis indicates that the threshold of detection in terms of planet temperature is very comparable to the one achieved with SPHERE, about 400K. As a consequence, the very same planets detected from the ground in the near IR with 8-m class telescopes will be potential targets for MIRI. MIRI, together with NIRCAM, will complete the spectral characterization and determine the energy balance of young giant planets in large-radius orbits. 

Since the coronagraphic suite of {\it JWST} was conceived (before 2005), several new developments have occurred in the field of exoplanet direct imaging, both at the instrumental level but more importantly at the data processing level. The MIRI performance will be enhanced by taking advantage of these new methods for data reduction that have shown significant improvement with respect to older techniques \citep[e.g.,][]{lafreniere2007, janson2008, soummer2012}.

\section{Acknowledgments}
The work presented is the effort of the entire MIRI team and the enthusiasm within the MIRI partnership is a significant factor in its success. MIRI draws on the scientific and technical expertise many organizations, as summarized in Papers I and II. 
A portion of this work was carried out at the Jet Propulsion Laboratory, California Institute of Technology, under a contract with the National Aeronautics and Space Administration.

We would like to thank the following National and International
Funding Agencies for their support of the MIRI development: NASA; ESA;
Belgian Science Policy Office; Centre Nationale D'Etudes Spatiales;
Danish National Space Centre; Deutsches Zentrum fur Luft-und Raumfahrt
(DLR); Enterprise Ireland; Ministerio De Economi{\'a} y Competividad;
Netherlands Research School for Astronomy (NOVA); Netherlands
Organisation for Scientific Research (NWO); Science and Technology Facilities
Council; Swiss Space Office; Swedish National Space Board; UK Space
Agency.

We also thank Mark Clampin for his many efforts to optimize the observatory on behalf of the coronagraph performance.


\eject

\begin{deluxetable}{lcccccc}
\tabletypesize{\scriptsize}
\tablecaption{Coronagraphic modes of MIRI}
\tablewidth{0pt}
\tablehead{
\colhead{Filter} & \colhead{Coronagraph} & \colhead{Stop transmission} & \colhead{Central wavelength} & \colhead{Bandwidth} & \colhead{IWA} & \colhead{Rejection}\\
		&		& [\%]		&	[$\mu m$]	& [$\mu m$]	& [arcsec]	& on-axis
}
\startdata
F1065C	& 4QPM1			& 62		& 10.575 	& 0.75	& 0.33	&  260 \\
F1140C	& 4QPM2			& 62		& 11.30 	& 0.8		& 0.36	&  285 \\
F1550C	& 4QPM3			& 62		& 15.50 	& 0.9		& 0.49	&  310\\
F2300C	& Lyot spot		& 72		& 22.75 	& 5.5 	& 2.16	&  850 \\
\enddata
\label{tab:filters}
\end{deluxetable}
\begin{deluxetable}{lcccccc}
\tabletypesize{\scriptsize}
\tablecaption{HR\,8799 planet's contrast in difference of magnitude.}
\tablewidth{0pt}
\tablehead{
\colhead{planet} & \colhead{F0560W} & \colhead{F0770W} & \colhead{F1000W} & \colhead{F1065C} & \colhead{F1140C} & \colhead{F1550C}\\
}
\startdata
b	& 10.95	& 10.85	& 9.60	& 9.73	& 9.16	& 9.02 \\
c, d	& 10.29	& 10.24	& 9.11	& 9.12	& 8.82	& 8.71
\enddata
\label{tab:hr8799}
\end{deluxetable}
\begin{deluxetable}{lcccccc}
\tabletypesize{\scriptsize}
\tablecaption{HR\,8799 planet's signal to noise ratio.}
\tablewidth{0pt}
\tablehead{
\colhead{planet} & \colhead{F1065C} & \colhead{F1140C} & \colhead{F1550C}\\
}
\startdata
b	& 17		& 26	& 11 \\
c	& 8		& 8	& 4 \\
d	& 2		& 4	& 6	
\enddata
\label{tab:hr8799snr}
\end{deluxetable}
\begin{deluxetable}{lcccccc}
\tabletypesize{\scriptsize}
\tablecaption{Proposed non-standard coronagraphic modes.}
\tablewidth{0pt}
\tablehead{
\colhead{Filter} & \colhead{Coronagraph} & \colhead{Stop transmission} & \colhead{Central wavelength} & \colhead{Bandwidth} & \colhead{IWA} & \colhead{Rejection}\\
		&		& [\%]		&	[$\mu m$]	& [$\mu m$]	& [arcsec]	& on-axis
}
\startdata
F0560W	& Lyot bar			& 100	& 5.60	& 1.15 	& 0.72	&  210 \\
F0770W	& Lyot bar			& 100	& 7.65	& 2.1 	& 0.72	&  55 \\
F1000W	& Lyot bar			& 100	& 9.95	& 1.9 	& 0.72	&  50 \\
\enddata
\label{tab:filters2}
\end{deluxetable}


\clearpage

\clearpage

\clearpage

\clearpage

\begin{figure}[t]
\centerline{
\includegraphics[width=8cm, angle=0]{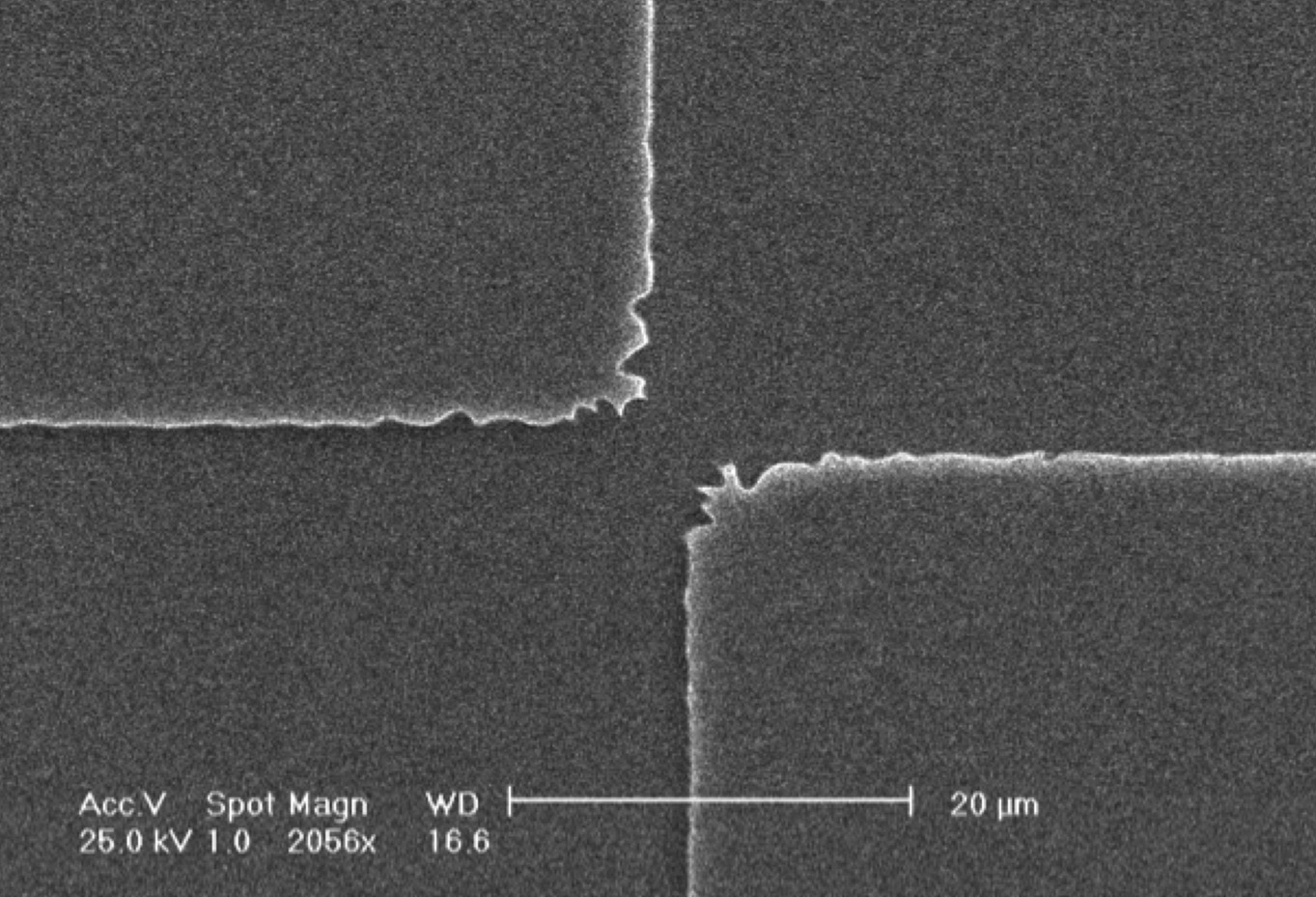}}
\caption{Microscopic photography of the one of the 4QPMs manufactured at CEA (Commissariat \'a l'Energie Atomique, France). The central defect is about 7$\mu m$ in width. As a comparison, the size of the JWST PSF  in the focal plane at 11.40$\mu m$  is $\lambda$F/D=230$\mu m$.
 }
\label{fig:pupil}
\end{figure}

\clearpage

\begin{figure}[t]
\centerline{
\includegraphics[width=15cm]{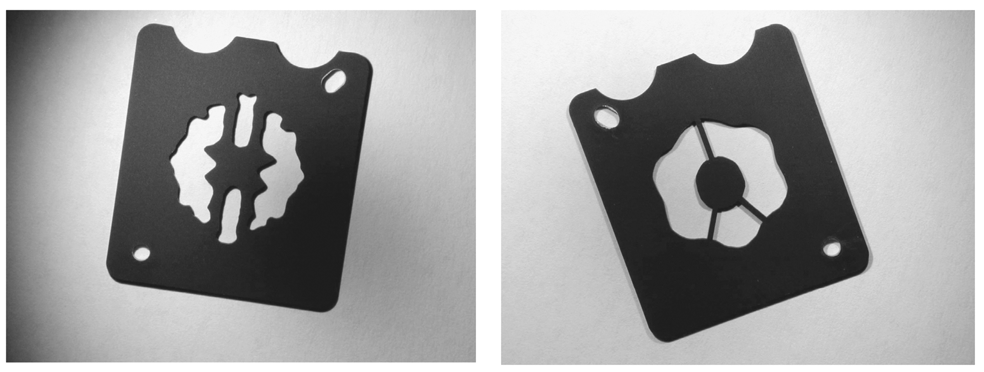}}
\caption{Coronagraph pupil masks, manufactured at LESIA (the Laboratory for Space Studies and Asrtrophysics 
Instrumentation, at the Paris Observatory). The 4QPM version is to the left and the Lyot one to the right.}
\label{fig:pupil}
\end{figure}


\clearpage

\begin{figure}[t]
\centerline{
\includegraphics[width=10cm]{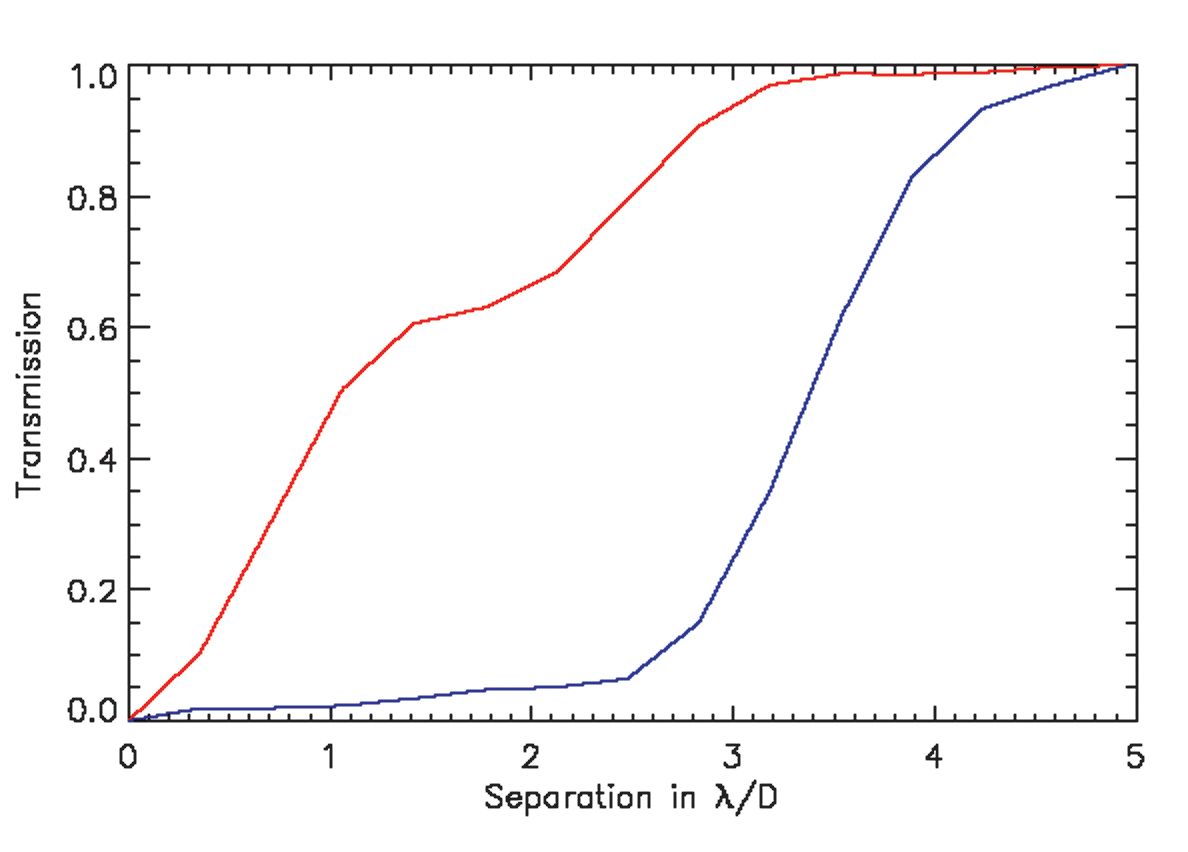}}
\caption{Radial transmission of the MIRI coronagraphs (4QPM in red, Lyot spot in blue).}
\label{fig:iwa}
\end{figure}


\clearpage

\begin{figure}[t]
\centerline{
\includegraphics[width=10cm]{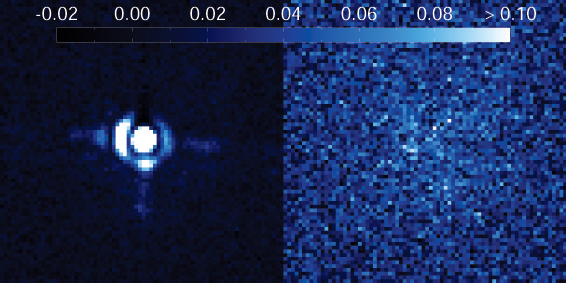}}
\caption{Sample images at 11.40 $\mu$m: left without the coronagraph, right with it. The field of view is 8.36" (76 detector pixels) and the color scale indicates the contrast with respect to the PSF maximum. The level of the coronagraphic image is close to the background noise and the central residual is attributed to the size of the source, which is not perfectly point-like. }
\label{fig:fmimage}
\end{figure}


\clearpage

\begin{figure}[t]
\centerline{
\includegraphics[width=10cm]{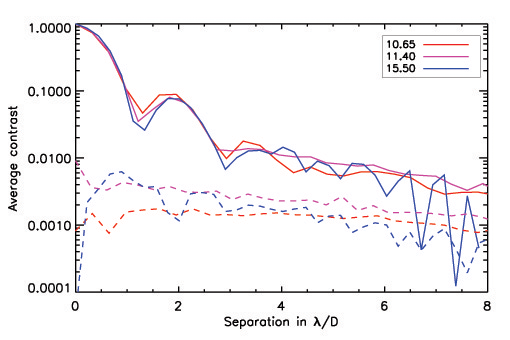}}
\caption{Contrast achieved in the test of the flight 4QPM coronagraphs (red for 10.65  $\mu$m, magenta for 11.40 $\mu$m and blue for 15.50  $\mu$m) in the flight imager. The solid lines are for  the unattenuated PSF and the dashed ones show the effect of the 4QPM.}
\label{fig:fmcontrast}
\end{figure}


\clearpage

\begin{figure}[t]
\centerline{
\includegraphics[width=8cm]{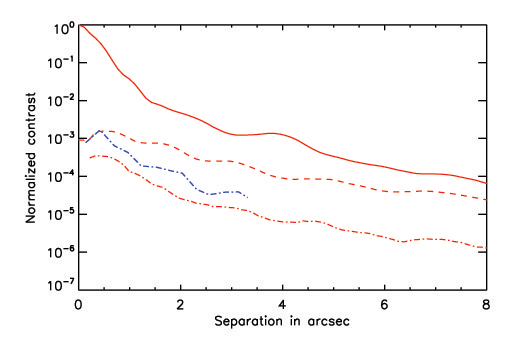}
\includegraphics[width=8cm]{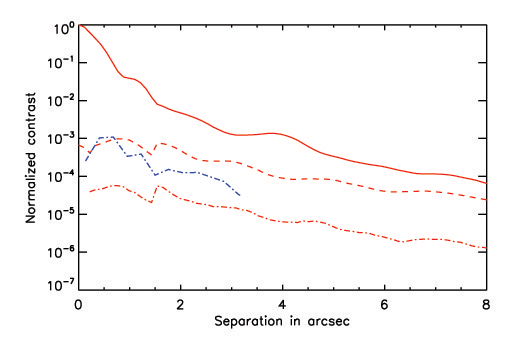}}
\caption{Normalized noise-free contrasts obtained under Case A (smaller WFE, in red)  and case B (larger WFE, in blue: only the reference star subtracted final result) in F1140C (left) and F1550C (right) filters on the PSF (solid) and the raw coronagraphic image (dashed). Estimated 5$\sigma$ contrasts using reference star subtraction are also shown (dash-dotted).}
\label{fig:rawcontrast}
\end{figure}

\clearpage

%


\clearpage

\begin{figure}[t]
\centerline{
\includegraphics[width=9.cm]{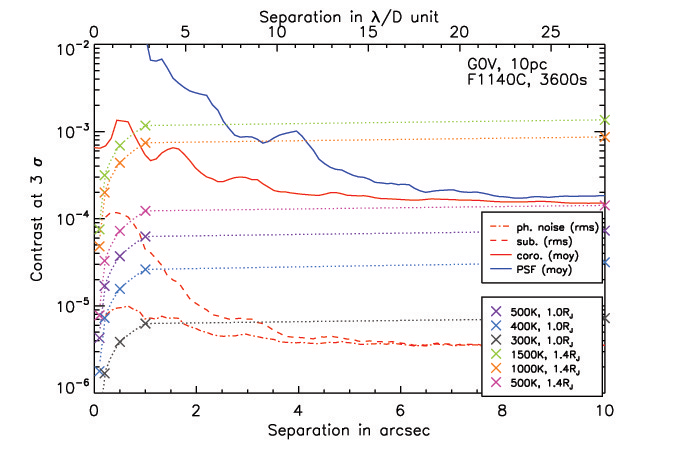}\hspace{-0.75cm}
\includegraphics[width=9.cm]{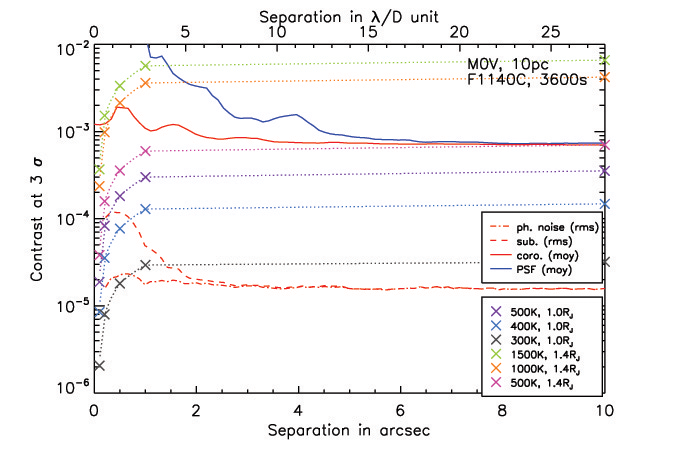}}
\centerline{
\includegraphics[width=9.cm]{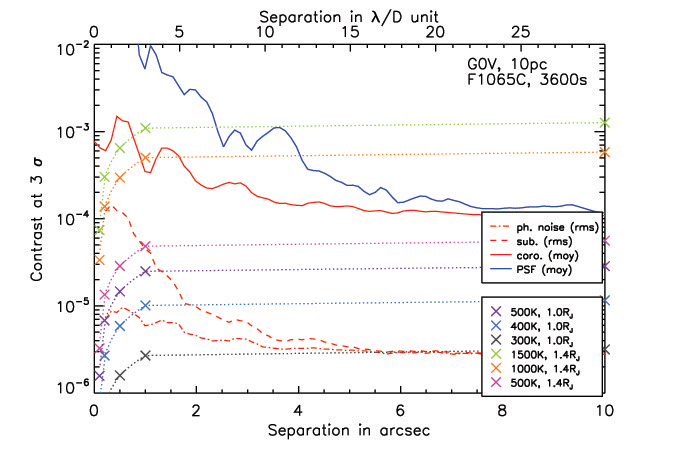}\hspace{-0.75cm}
\includegraphics[width=9.cm]{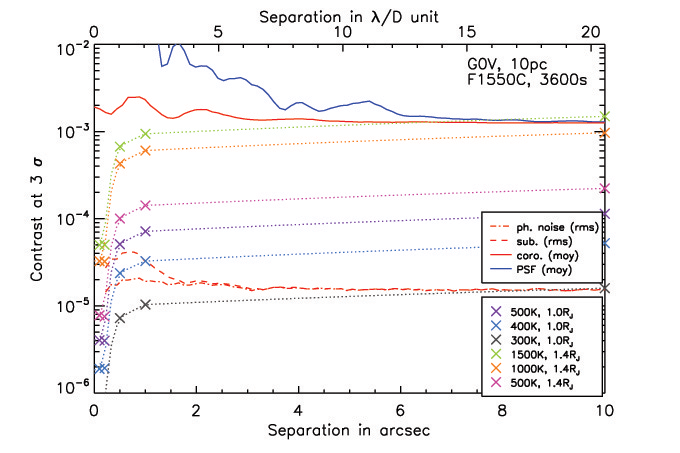}}
\caption{Estimated performance of MIRI in a few selected cases, assuming Case A (smaller WFE) and including the effects of noise. See text for details.}
\label{fig:perf}
\end{figure}


\clearpage

\begin{figure*}
\centerline{\includegraphics[width=15cm]{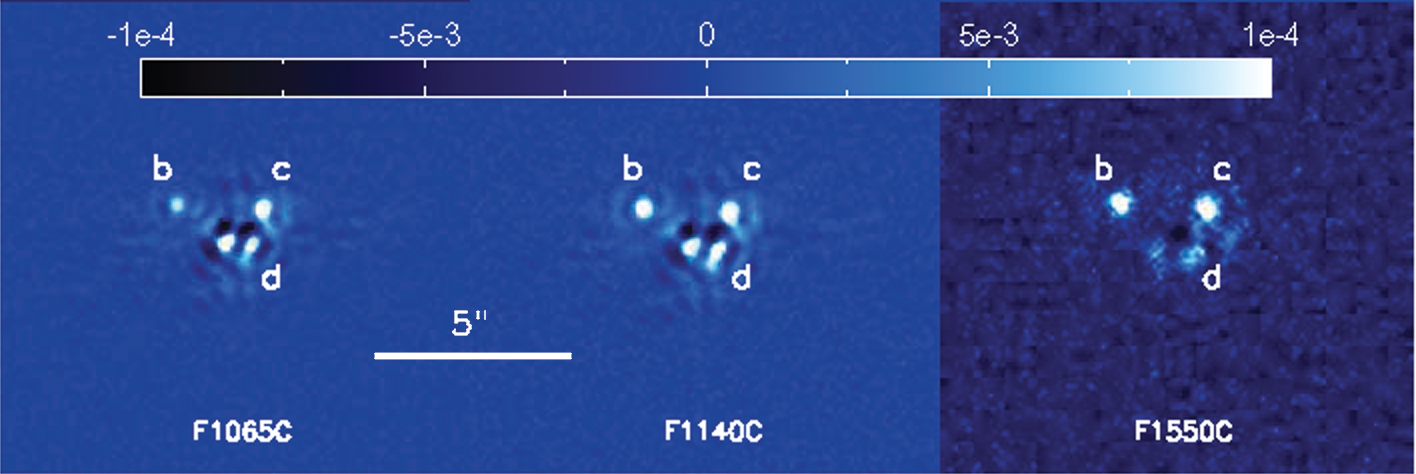}}
\caption{Simulated observations of the HR\,8799 system in the 4QPM coronagraphic filters, assuming Case A (smaller WFE). The planets are labeled b, c and d. The image is in units of contrast relative to the un-occulted star, as displayed by the color scale.}
\label{fig:imageshr8799}
\end{figure*}

\clearpage

\begin{figure}
\centerline{\includegraphics[width=12cm]{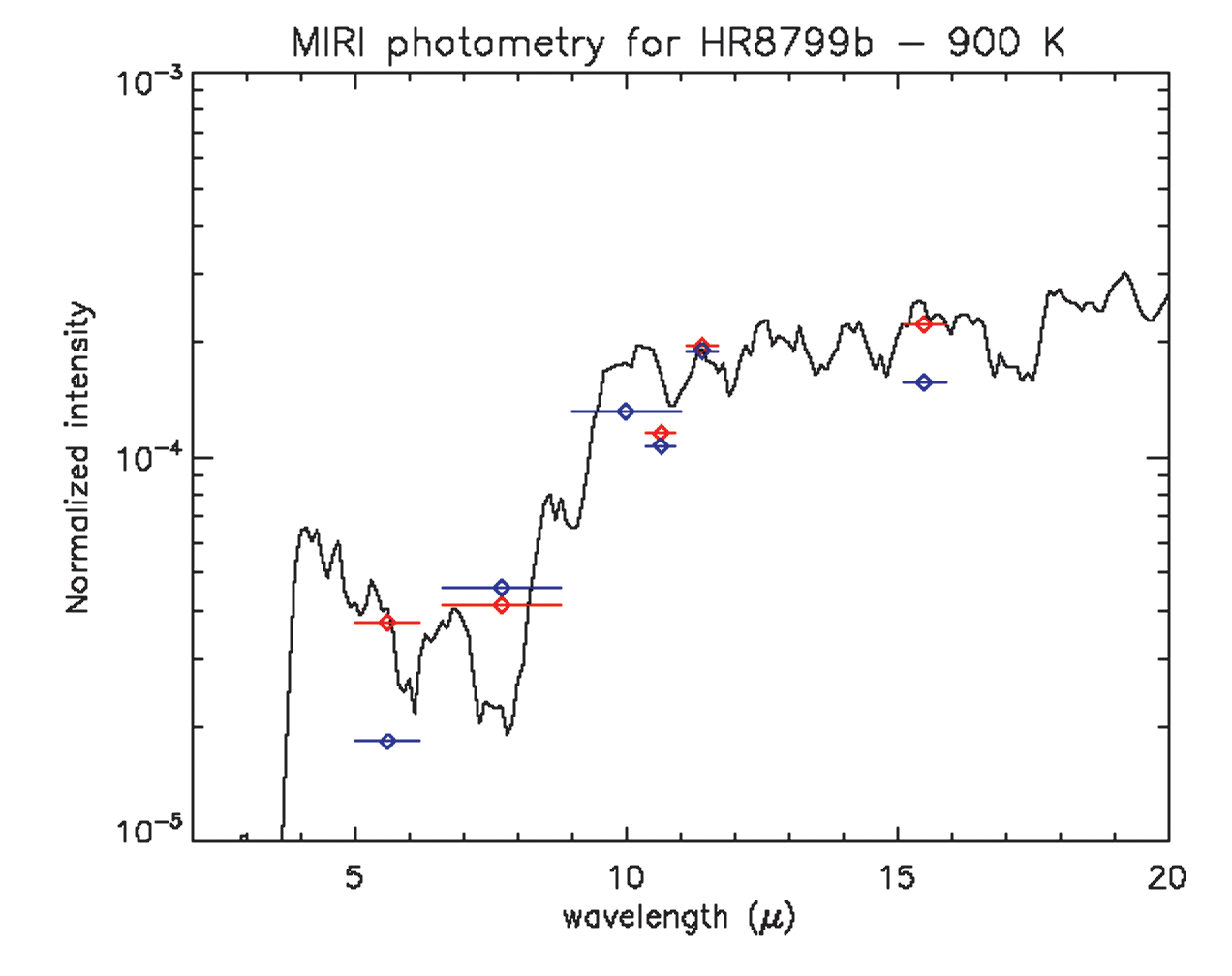}}
\caption{Spectral Energy distribution (expressed in contrast) of the planet HR8799 b, assuming Case A (smaller WFE). The solid line represents the spectral model \citep{allard2001} degraded to a resolution of $R=20$. The red and blue points represent the true and measured intensities respectively in several MIRI filters (F560W, F770W, F1000W, F1065C, F1140C, F1550C). To get high quality measurements in the first three filters may require implementation of the coronagraph bar observations discussed in Section 7. Horizontal bars indicate the width of filters.}
\label{fig:spectrumhr8799}
\end{figure}

\clearpage

\begin{figure*}
\centerline{\includegraphics[width=15cm]{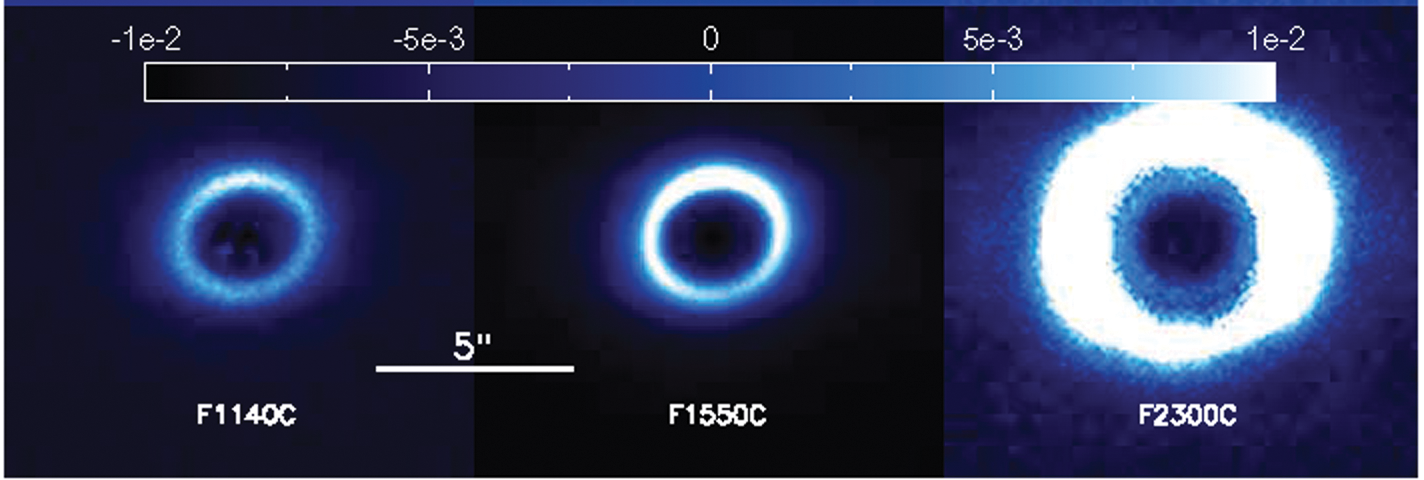}}
\caption{Simulated observations of the HD\,181327 debris disk  in the coronagraphic filters, assuming Case A (smaller WFE). The image is in units of contrast relative to the un-occulted star, as displayed by the color scale. The cross effect from the 4QPM is not included in this simulation.}
\label{fig:hd181327}
\end{figure*}

\clearpage

\begin{figure*}
\centerline{
\includegraphics[width=9.cm]{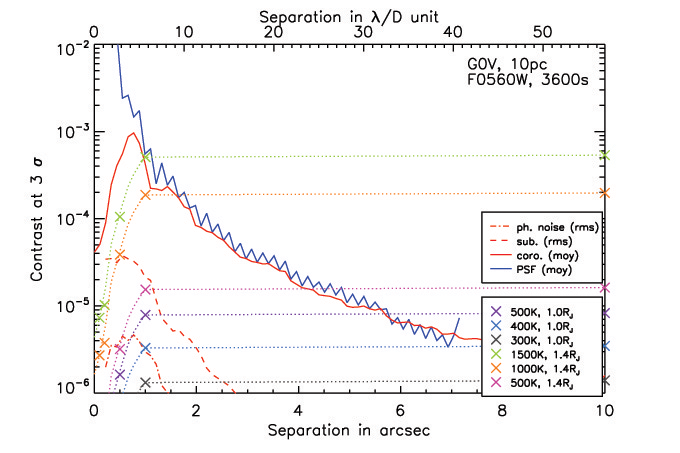}\hspace{-0.75cm}
\includegraphics[width=9.cm]{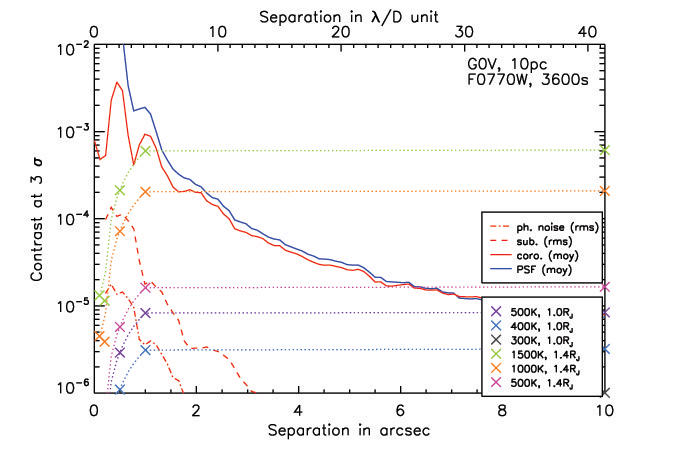}}
\caption{Same as  Figure \ref{fig:perf} for the Lyot bar mask and assuming Case A (smaller WFE)}
\label{fig:perf2}
\end{figure*}

\end{document}